\documentclass[reprint, amsmath,amssymb, aps, showpacs, superscriptaddress, pra]{revtex4-1}

\usepackage{graphicx}
\usepackage{tensor}
\usepackage{verbatim}
\usepackage[breaklinks=true,colorlinks=true,linkcolor=blue,urlcolor=blue,citecolor=blue]{hyperref}
\def\expect#1{\left\langle #1 \right\rangle}
\def\bra#1{{\langle #1 |}}
\def\ket#1{{| #1 \rangle}}
\newtheorem{definition}{Definition}
\newtheorem{prop}{Proposition}
\newtheorem{lemma}{Lemma}

\begin{document}
\preprint{}

\title{Low depth measurement-based quantum computation beyond two-level systems}
\author{Timothy J. Proctor}
 \email{tjproctor@berkeley.edu}
\affiliation{School of Physics and Astronomy, E C Stoner Building, University of Leeds, Leeds, LS2 9JT, UK}
\affiliation{Berkeley Quantum Information and Computation Center, Department of Chemistry, University of California, Berkeley, CA 94720, USA}
\date{\today}

\begin{abstract}
Low depth measurement-based quantum computation with qudits ($d$-level systems) is investigated and a precise relationship between this powerful model and qudit quantum circuits is derived in terms of computational depth and size complexity. To facilitate this investigation a qudit `unbounded fan-out' circuit model, in which a qudit may be quantum-copied into an arbitrary number of ancillas in a single time-step, is introduced and shown to be capable of implementing interesting $n$-qudit unitaries in constant depth. A procedure for reducing the quantum computational depth in the measurement-based model is then proposed and using this it is then shown that there is a logarithmic depth separation between the depth complexity of qudit measurement-based computation and circuits composed of gates act on a bounded number of qudits. The relationship is made precise by showing that the depth complexity of the qudit measurement-based model is exactly equivalent to that of unbounded fan-out circuits. These results illustrate that the well-known advantages inherent in qubit measurement-based quantum computation are also applicable to the higher-dimensional generalisation. As qudits are both naturally available and have been shown to provide fundamental advantages over binary logic encodings, this then suggests that the qudit measurement-based model is a particularly appealing paradigm for universal quantum computation.
\end{abstract}

\pacs{}

\maketitle

\section{Introduction}
Most physical quantum systems are naturally many-levelled and consequently there is no fundamental reason to restrict the development of quantum information processing protocols to qubit-based binary logic. Indeed, although the main focus in the literature has been on qubits, there is growing evidence that there are intrinsic advantages to harnessing these naturally available higher-dimensional systems for quantum information processing \cite{sheridan2010security,lanyon2009simplifying,parasa2011quantum,zilic2007scaling,parasa2012quantum,campbell2014enhanced,anwar2014fast,campbell2012magic,andrist2015error,duclos2013kitaev}. In the specific context of quantum computation, a particularly striking motivation for qudit-based logic is that increasing the dimension of the computational systems improves both the robustness of algorithms \cite{parasa2011quantum,zilic2007scaling,parasa2012quantum} and fault-tolerance thresholds \cite{campbell2014enhanced,anwar2014fast,campbell2012magic,andrist2015error,duclos2013kitaev}. In addition to these advantages, further encouragement to extend such investigations into higher-dimensional protocols is provided by the impressive range of experiments demonstrating quantum control of qudits \cite{bent2015experimental,smith2013quantum,walborn2006quantum,neeley2009emulation,lima2011experimental,rossi2009multipath,dada2011experimental}. 
\newline
\indent
It has been known since Raussendorf and Briegel introduced the \emph{one-way quantum computer} \cite{raussendorf2001one} that adaptive local measurements of \emph{qubits} prepared in an entangled state are sufficient for universal quantum computation. This remarkable computational paradigm is particularly appealing from a physical perspective as it allows the creation of entanglement to be separated into an initial off-line procedure. Indeed, some of the most promising demonstrations of the basic building blocks required for a physical quantum computer have been implementations of this computational model \cite{lanyon2013measurement,bell2014experimental,tame2014experimental,chen2007experimental}. That the one-way quantum computer is universal is perhaps initially a surprise as on the surface it may appear to have very little in common with the quantum circuit model. However, the relationship between the two models has been extensively researched and is now well understood \cite{raussendorf2003measurement,danos2007measurement,Broadbent20092489,danos2009extended,browne2011computational}, with the interesting conclusion that the one-way model requires less (quantum) computational steps to implement certain operator sequences than ordinary quantum circuits \cite{Broadbent20092489}. Specifically, Browne \emph{et al.} \cite{browne2011computational} showed that the one-way model has exactly the same computational complexity as a quantum circuit model in which a gate that quantum-copies a qubit into an unbounded number of ancillas may be implemented in a single time-step, known as the \emph{unbounded fan-out model} and first investigated in detail by H\o yer and \v{S}palek \cite{hoyer2005quantum,hoyer2003quantum}. These results highlight that the one-way model is not only physically appealing but also has fundamental advantages over quantum computation using unitary gates alone.
\newline
\indent
Higher-dimensional systems and the one-way model are both particularly promising paradigms for quantum computation and hence it is interesting that one-way  computation may be formulated with qudits, as shown by Zhou \emph{et al.} \cite{zhou2003quantum}. However, in contrast to the binary case, the relationship between this model and qudit circuits has not been addressed and this is the subject of this paper. To the knowledge of this author, the qudit quantum circuits which will be necessary to make precise this relationship have not previously been defined and investigated, and hence this is first considered in Section~\ref{circuits}. This includes the introduction of a qudit unbounded fan-out model, which itself may be of independent interest. The qudit one-way model is then presented in Section~\ref{dpatterns}, in which a generalisation of the \emph{standardisation} procedure of Danos \emph{et al.} \cite{danos2007measurement} is introduced. This will be used to investigate low depth one-way computations and show that there is a logarithmic depth separation between the qudit one-way model and quantum circuits containing gates acting on a fixed number of qudits. The relationship is then made precise by showing that the qudit unbounded fan-out and one-way models have exactly the same depth complexity, generalising the result of Ref.~\cite{browne2011computational} to multi-valued logic. Finally, a constructive method for implementing any Clifford circuit in constant depth will be given. These results confirm that the qudit one-way model is a particularly promising paradigm for quantum computation, exhibiting both the advantages of quantum multi-valued logic and of hybrid quantum-classical processing. To begin, the relevant formalism for higher-dimensional quantum computation is introduced in Section~\ref{dintro}.

\section{Quantum computation with qudits \label{dintro}}
The quantum systems of interest have a Hilbert space of some arbitrary finite dimension $d \in \mathbb{N}$ with $d \geq 2$. For typographical simplicity the dimension $d$ will be suppressed in all of the following notation and it is to be assumed that all objects (i.e., operators etc) are defined on arbitrary dimension qudits (all of the same dimension).
\subsection{The computational and conjugate bases}
 An orthonormal \emph{computational basis} may be arbitrarily chosen and denoted
 \begin{equation} \mathcal{B}:=\{\ket{n} \mid n \in \mathbb{Z}(d) \} ,\end{equation} where $\mathbb{Z}(d)= \{0,1,...,d-1\}$. A \emph{conjugate basis}, may be defined in terms of this basis by
   \begin{equation} \mathcal{B}_{+} :=\{\ket{+_n} := F \ket{n} \mid n \in \mathbb{Z}(d) \}, \end{equation}
where $F$ is a unitary Fourier transform operator defined by
\begin{equation} F\ket{n}:= \frac{1}{\sqrt{d}} \sum_{m=0}^{d-1} \omega^{mn} \ket{m},\end{equation}
with $\omega = \exp(2 \pi i /d)$ the $d\textsuperscript{th}$ root of unity. For the special case of a qubit ($d=2$) the Fourier transform reduces to the well-known Hadamard gate. An important property of $F$ is that it has order $4$, i.e., $F^4 = \mathbb{I}$ \cite{vourdas2004quantum}. It is simple to confirm that
   \begin{equation} \expect{m,+_n} = \frac{ \omega^{mn}}{\sqrt{d}}  \hspace{0.5cm} \forall m,n \in \mathbb{Z}(d), \label{mub} \end{equation}
and hence the bases are mutually unbiased \cite{durt2010mutually,noteMBQC_A}.
The (generalised) \emph{Pauli operators} are the unitaries defined by
\begin{equation} Z \ket{n}  := \omega^n \ket{n} , \hspace{1cm} X \ket{n}  := \ket{n+1} ,  \end{equation}
where the arithmetic is modulo $d$ and this is to be assumed for all arithmetic in the following unless otherwise stated. For qubits these unitaries reduce to the ordinary $U(2)$ Pauli operators. 
It may be easily confirmed that their action on the conjugate basis is
\begin{equation} X \ket{+_n} = \omega^{-n} \ket{+_n} ,\hspace{0.5cm}  Z \ket{n} = \ket{+_{n+1}} \label{basisshift},\end{equation}
and hence the computational and conjugate bases are eigenstates of $Z$ and $X$ respectively. The Pauli operators obey the Weyl commutation relation
\begin{equation} Z^a X^b = \omega^{ab} X^b Z^a,\end{equation}
with $a,b \in \mathbb{Z}(d)$.

\subsection{Universal sets of unitaries}
Any two-qudit entangling unitary along with a set of single-qudit unitaries that can generate any single-qudit unitary may be used to generate all of $SU(d^n)$ \cite{brylinski2002universal}. The canonical entangling unitaries in quantum computation are the controlled gates and the controlled-$Z$ and controlled-$X$ gates are  defined by
\begin{align} CZ \ket{m}\ket{n}&:= \omega^{mn} \ket{m}\ket{n} ,\\ 
 CX \ket{m}\ket{n}&:=  \ket{m}\ket{m+n} ,
\end{align}
respectively. When necessary, super and subscripts will be used to denote the control and target qudits respectively, i.e., $C^j_kX\ket{m}_j\ket{n}_k=\ket{m}_j\ket{m+n}_k$.
An important class of single-qudit unitaries are the rotation gates. These operators take a vector parameter $\boldsymbol{\theta} = (\theta_0,\theta_1,...,\theta_{d-1}) \in \mathbb{R}^{d}$ and are defined by
\begin{equation} R(\boldsymbol{\theta})  \ket{n} := e^{ i \theta_n} \ket{n} .\label{Zdt}\end{equation}
Then taking $ v(\boldsymbol{\theta}) : = FR(\boldsymbol{\theta})$, Zhou \emph{et al.} \cite{zhou2003quantum} have shown that the set of all such operators may exactly generate any single-qudit unitary \cite{noteMBQC_B}. Hence a universal set of unitaries for qudits is
\begin{equation}   \mathcal{G}_{\text{uni}}: = \{ CZ, v(\boldsymbol{\theta})  \mid  \boldsymbol{\theta} \in \mathbb{R}^d \} , \label{Euni} \end{equation}
 and this will be important herein. This includes a continuum of single-qudit operators, however the set $\mathcal{G}_{\text{uni}}: = \{ CZ, F, v(\boldsymbol{\theta}) \})$ for a `generic' fixed $\boldsymbol{\theta}$ generates a dense subset of $SU(n^d)$ \cite{Proctor2015ancilla}, and hence can approximate any unitary to any desired accuracy.

  \subsection{The Pauli and Clifford groups}
  The Pauli and Clifford groups play a fundamental role in the one-way model. The single-qudit Pauli group is defined here to be
\begin{equation} \mathcal{P}_1 :  = \{p_{\xi ,a, b} = \widehat{\omega}^{\xi} X^a Z^b \mid  \xi \in \mathbb{Z}(D), \hspace{0.1cm} a,b \in \mathbb{Z}(d) \}, \end{equation} 
where $\widehat{\omega}$ is the $D^{th}$ root of unity and  \cite{de2013linearized,farinholt2014ideal,noteADQC1}
\begin{equation} D: = \left\{ 
  \begin{array}{l l}
    d & \quad \text{for odd $d$}\\
    2d & \quad \text{for even $d$.}
  \end{array} \right. \end{equation}
  
The $n$-qudit Pauli group, denoted $\mathcal{P}$, is the subset of $U(d^n)$ consisting of operators of the form
\begin{equation} p_{\xi, \boldsymbol{v}} : = p_{\xi_1 a_1b_1} \otimes p_{\xi_2 a_2b_2} \otimes .... \otimes p_{\xi_n a_nb_n} ,\end{equation}
where $\xi= \xi_1+\xi_2+...+\xi_n$ with the addition modulo $D$ and $\boldsymbol{v}=(a_1,...,a_n,b_1,...,b_n) \in \mathbb{Z}(d)^{2n}$. The ($n$-qudit) Clifford group is the normaliser of this group in $U(d^n)$ and hence is defined by
\begin{equation} \mathcal{C}: = \{ U \in U(d^n) \mid U  p U^{\dagger} \in \mathcal{P} \hspace{0.2cm} \forall   p  \in \mathcal{P} \}. \end{equation}
Gates from this set alone are not sufficient for universal quantum computation and furthermore (a generalisation of) the Gottesman-Knill theorem shows that computations using gates from this set along with only qudits measured and prepared in Pauli eigenstates may be efficiently classically simulated \cite{hostens2005stabilizer,de2013linearized,van2013efficient,gottesman1999fault}. Recently Farinholt \cite{farinholt2014ideal} has given a minimal set of generators for the Clifford group for arbitrary dimension $d$ (common constructions, such as those in \cite{gottesman1999fault,hall2005cluster}, apply only to prime $d$). The results in Ref.  \cite{farinholt2014ideal} imply that a minimal generating set is given by
\begin{equation}\mathcal{C} = \langle F, P, CZ \rangle, \label{Cliffgen} \end{equation} 
where $P$ is the \emph{phase gate} defined by 
\begin{equation} P \ket{n}:= \omega^{\frac{n}{2}(n+\delta_d)} \ket{n}, \end{equation}
with $\delta_d=0$ ($\delta_d=1$) for $d$ even (odd) and it is noted that the arithmetic in this definition is not modulo $d$. For $d=2$ this reduces to the well-known qubit phase gate $P= \ket{0}\bra{0}+i \ket{1} \bra{1}$. Straightforward algebra may be used to confirm that the conjugation relations of these generators on arbitrary Pauli operators are
\begin{align}  F  p_{\xi ,a,b}   F^{\dagger}&= p_{\xi+\xi_F(ab),-b,a}, \label{DcliffF} \\
  P  p_{\xi ,a,b} P^{\dagger} &= p_{\xi+\xi_P(a) ,a,a+b} , \label{DcliffP} \\ 
 CZ p_{\xi ,(a_1 ,a_2 ,b_1,b_2)} C Z^{\dagger} & =    p_{\xi ,(a_1,a_2,b_1+a_2,b_2+a_1)} \label{DcliffCZ}, \end{align}
where the changes in the phase factors are given by
\begin{align}
\xi_F(n)&=n(\delta_d-2), \\
\xi_P(n)&=n(1-(n-1)(\delta_d-2)/2), \label{Pconjphase}
\end{align}
and are often (but not always) of no importance.

\subsection{Quantum computation}
It is convenient to introduce the ideas of quantum computation in a model-independent fashion which may then be applied to both quantum circuits and the one-way model. Define a (qudit) \emph{quantum computational model} by $\mathfrak{M} =(\mathcal{A},\mathcal{S})$ where $\mathcal{A}$ is a set of basic allowed operations (which act on qudits) and $\mathcal{S}$ is some set of preparable states. Operations may in general have classical outputs or depend on classical inputs (e.g., measurement outcomes).  A \emph{quantum computation} in a particular model is then a quadruplet $\mathfrak{Q} = ( \mathcal{V},\mathcal{I},\mathcal{O},\mathfrak{q} )$ where $ \mathcal{V}$ is a set of qudits, $\mathcal{I},\mathcal{O} \subseteq \mathcal{V}$ are \emph{input} and \emph{output} subsets and $\mathfrak{q}$ is a sequence of \emph{operations} from $\mathcal{A}$ which act on qudits from $ \mathcal{V}$. Operations may only depend on outputs from operations earlier in the operation sequence. All non-input states $ \mathcal{V} \setminus \mathcal{I}$ are prepared in states from $\mathcal{S}$ and it is assumed that the input qudits may in general be in an arbitrary state $\ket{\psi}$.  A quantum computation may be considered to implement the unitary operation $U \in U(d^{|\mathcal{I}|})$ if for the any input state $\ket{\psi}$, the final state of the output qudits is $U\ket{\psi}$ (which requires $|\mathcal{O}|=|\mathcal{I}|$) and such a computation will be denoted $\mathfrak{Q}_U$. The model is considered to be universal if it may implement any unitary on any number of input qudits (or approximately universal if it can only approximate any unitary to arbitrary accuracy). Qudits that are not in the input or output sets are normally termed \emph{ancillas}.
\begin{definition}
For a quantum computation $\mathfrak{Q} = ( \mathcal{V},\mathcal{I},\mathcal{O}, \mathfrak{q} )$, a path of dependent operations is a sub-sequence $(q_j)$ of $\mathfrak{q}$ such that each operation either 
\newline
\indent
(a)  acts on a qudit in common with, or
\newline
\indent
(b) depends upon the outcome of,
\newline
the previous operation in the sub-sequence.
\end{definition}
Using this the (quantum) depth of a computation may be defined.
\begin{definition}
The quantum depth of a quantum computation $\mathfrak{Q}$, denoted $\mathrm{depth}(\mathfrak{Q})$, is the number of operations in the longest path of dependent operations. 
\end{definition}
The depth represents the number of steps required for the computation and hence such a definition of depth encodes the idea that two operations cannot be performed simultaneously on a qudit and that an operation may not be performed simultaneously with one whose output it depends upon. The quantum size of a quantum computation, denoted $\mathrm{size}(\mathfrak{Q})$, is the sum of the size of each operation it contains where the size of an operation is defined to be the number of qudits on which it acts. Note that these are referred to as \emph{quantum} size and depth as they take no account of any classical computational resources required for any manipulations of any classical outputs. However, these quantities are clearly physically motivated given the relative practical difficulty of classical and quantum computation \cite{noteMBQC_C}. 
\newline
\indent
For two computations $\mathfrak{Q}_0$ and $\mathfrak{Q}_1$ such that $\mathcal{O}_0=\mathcal{I}_1$ (which may be enforced with a qudit relabelling as long as $|\mathcal{O}_0|=|\mathcal{I}_1|$) the composite `serial' computation  $\mathfrak{Q}_1 \mathfrak{Q}_0$ may be defined in a natural way as 
\begin{equation} \mathfrak{Q}_1 \mathfrak{Q}_0 :=(\mathcal{V}_0 \cup \mathcal{V}_1,\mathcal{I}_0,\mathcal{O}_1,\mathfrak{q}_1\mathfrak{q}_0), \end{equation}
where $\mathfrak{q}_1\mathfrak{q}_0$ is the concatenated operation sequence (the $\mathfrak{q}_0$ command sequence followed by the $\mathfrak{q}_1$ command sequence). In a similar way for $\mathcal{V}_0 \cap \mathcal{V}_1 = \emptyset$, the `parallel' tensor product of two computations may defined by 
\begin{equation} \mathfrak{Q}_1\otimes \mathfrak{Q}_0 :=(\mathcal{V}_0 \cup \mathcal{V}_1,\mathcal{I}_0 \cup \mathcal{I}_1 ,\mathcal{O}_1 \cup \mathcal{O}_1,\mathfrak{q}_1\mathfrak{q}_0). \end{equation}
It is easily confirmed that $\text{depth}(\mathfrak{Q}_1 \mathfrak{Q}_0) \leq  \text{depth}(\mathfrak{Q}_1)+\text{depth}( \mathfrak{Q}_0)$ and $\text{depth}(\mathfrak{Q}_1 \otimes \mathfrak{Q}_0) = \text{max} ( \text{depth}(\mathfrak{Q}_1),\text{depth}( \mathfrak{Q}_0))$ and in both cases sizes add.

\section{qudit quantum circuits \label{circuits}}
A qudit quantum circuit model (QCM) is any quantum computational model in which the allowed operations are some set of unitary operators (quantum gates). It is conventional to restrict the set of preparable states to $ \mathcal{S}=\{\ket{0}\}$ and this will be taken to be the case herein.
\subsection{Standard and unbounded fan-out circuits}
\begin{figure}
\includegraphics[width=8.5cm]{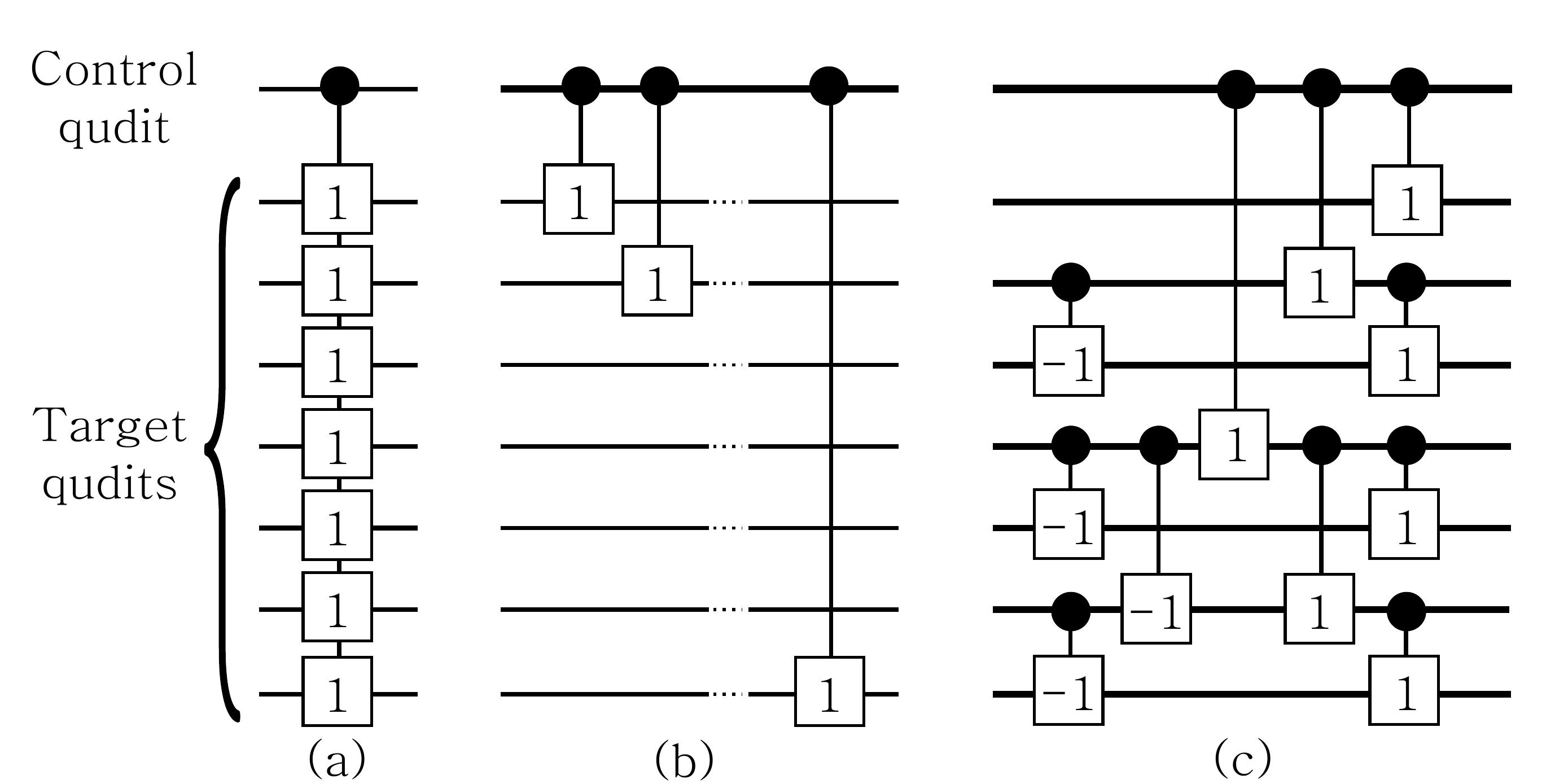}
\caption{\label{quditfanout}(a) A circuit notation for the $n$-qudit fan-out gate. (b) The fan-out gate decomposition into $n$ controlled-$X$ gates with a circuit size and depth of $O(n)$. This uses the notation that a square box containing an $n$ denotes $X^n$. (c) An alternative circuit decomposition implementing the qudit fan-out gate with a depth of $O( \log n )$ and size $O(n)$. The case shown here is for $n=2^3-1$. The same structure may be used for all $n=2^m-1$ and for other cases the structure for $2^{\lceil \log (n+1) \rceil}-1$ may be used (with certain gates omitted).}
\end{figure}

\begin{definition} The `standard quantum circuit model' is a QCM in which the allowed gate set is some universal set containing only gates that act on a fixed (i.e., independent of input size) number of qudits.
\end{definition}
A `standard quantum circuit' is then a particular computation in this model. The exact specification of the gate set is not required when considering only how computational depth and size scale with the number of input qudits for the implementation of $n$-qudit unitary families, and for concreteness the gate set may be taken to be $\mathcal{G}_{\text{uni}}$ (as defined in Eq.~(\ref{Euni})). This is because any $m$-qudit gate may be simulated exactly with $O(d^{2m})$ two-qudit gates \cite{bullock2005asymptotically}, which for constant $m$ is $O(1)$ with respect to input size $n$. These gates may in turn be decomposed into gates from $\mathcal{G}_{\text{uni}}$.
\newline
\indent
An alternative circuit model which it will be seen does not have the same depth complexity as standard quantum circuits can be defined by first introducing the $n$-qudit fan-out gate:
\begin{equation} \label{Efanout} \textsc{fanout} \ket{x} \ket{y_1,...,y_{n}}: = \ket{x}   \left| y_1+x,...,y_n+ x \right\rangle. \end{equation} 
For general $d$ the fan-out gate is not self-inverse and has order $d$. A circuit notation for this gate is defined in Fig.~\ref{quditfanout} (a) and it is obvious that this gate may be composed from a sequence of $n$ controlled-$X$ gates as shown in the circuit diagram of Fig.~\ref{quditfanout} (b). The gate is so named because it may be used to copy computational basis states into $n$ qudits, which may be achieved by setting all $y_j=0$ in Eq.~(\ref{Efanout}).

\begin{definition}
The `unbounded fan-out model' is a QCM is which the allowed gate set is some universal set containing gates that act on a fixed (i.e., independent of input size) number of qudits along with fan-out gates on any number of qudits.
\end{definition}
Again, an `unbounded fan-out circuit' is then a particular computation in this model.
For the same reasons as given above, for concreteness the gate set may be taken to be $\mathcal{G}_f=\mathcal{G}_{\text{uni}} \cup \{\textsc{fanout} \}$. The standard asymptotic notation that a function $f(n)$ is $\Omega(g(n))$ ($O(g(n))$) if $f(n) > C g(n)$ ($f(n) < C g(n)$) for all $n > n_0$ for some constants $C >0$ and $n_0 $ is used  in the following proposition.

\begin{prop}\label{logdepthfanout} Any standard quantum circuit for the $n$-qudit fan-out gate has a depth of $\Omega(\log n)$ and there is a standard quantum circuit for this gate with a depth of $O(\log n)$.
\end{prop}
The proof is identical to that for the qubit sub-case proven in Ref.~\cite{fang2006quantum}. A circuit for the $n$-qudit fan-out gate with a depth of $O(\log n)$ is presented in Fig~\ref{quditfanout}. All the output qudits of the fan-out gate depend on the state of the control qudit. With $l$ circuit layers composed of gates that act on at most $m$ qudits for some constant $m$, at most $m^l$ qudits can depend on the state of the control qudit. Hence for $n$ qudits to depend on the control qudit it is necessary for at least $l=\log_m n$ layers, and hence any circuit must have a depth of $\Omega(\log n)$.
\newline
\indent
This proposition shows that the ability to implement gates on an unbounded number of qudits in unit depth allows for lower depth circuits. This obviously implies the following complexity relation between standard and unbounded fan-out circuits:
\begin{lemma} Any $n$-qudit unbounded fan-out circuit $\mathfrak{F}$ may be implemented with a standard quantum circuit that has a size of $O(\mathrm{size}(\mathfrak{F}))$ and a depth of $O(\mathrm{depth}(\mathfrak{F})\log n)$.
\end{lemma}
The relations between the one-way model and the circuit model will be stated in terms of unbounded fan-out circuits and this lemma may be used to convert these into statements about standard quantum circuits.
\subsection{Constant depth unbounded fan-out circuits}
\begin{figure}
\includegraphics[width=8.0cm]{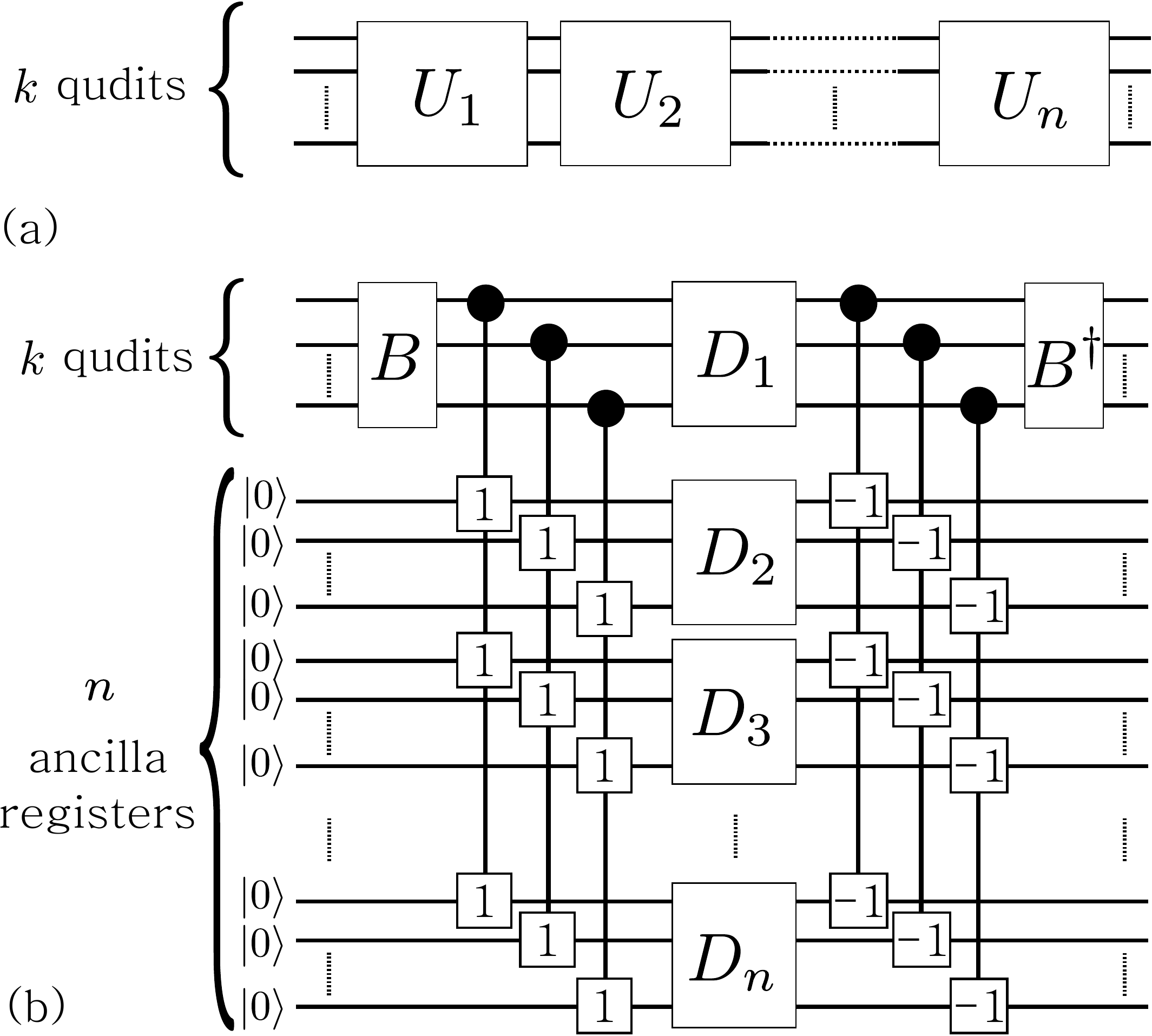}
\caption{\label{quditdiagonal}(a) A sequence of $n$ mutually commuting unitaries $U_j$ acting on a set of $k$ qudits. As these operators commute they are all diagonalised by some $k$-qudit unitary $B$. (b) A circuit which implements commuting gates in parallel where $D_j=BU_jB^{\dagger}$. This circuit requires $k(n-1)$ ancilla qudits. Each inverse fan-out gate may be implemented with $d-1$ fan-out gates.}
\end{figure}

Certain operators which may be implemented in constant depth with an unbounded fan-out circuit are now presented. Although these results are of independent interest, the main purpose of their presentation is that they will be required in later sections. The unbounded fan-out gate facilitates the application of commuting gates on a set of qudits at the same time whenever the basis in which they are all diagonal can be transformed into with a low depth circuit. More precisely:
\begin{prop} \label{diagprop} Consider a sequence of $n$ pair-wise commuting unitaries $U_i$ that act on $k$ qudits and which are hence diagonalised by the same operator $B$, i.e., $BU_iB^{\dagger}=D_i$ where for each $i$, $D_i$ is some diagonal unitary. Such an operator sequence may be implemented with an unbounded fan-out circuit with a depth of $\max_i (\text{depth} (D_i) )+ 2 \text{depth}(B) +d$ and a size of $\max ( O(n^2),O(\text{size}(B)),O \left(\sum \text{size}(D_i) \right) )$. 
\end{prop}
This generalises a result (for qubits) of Moore and Nilsson \cite{moore2001parallel}.
The proof is by way of a circuit diagram given in Fig.~\ref{quditdiagonal}.
This proposition will be of use later, but may also be applied to parallelise a variety of interesting quantum circuits (for example sequences of controlled rotations gates). Low and constant depth \emph{qubit} unbounded fan-out circuits have been investigated in detail by H\o yer and \v{S}palek \cite{hoyer2005quantum,hoyer2003quantum} and other authors \cite{takahashi2013collapse,takahashi2010quantum,moore2001parallel} and have been shown to be remarkably powerful. A large range of operations are implementable with such circuits, including an approximation of the quantum Fourier transform with arbitrary modulus  \cite{hoyer2005quantum} which is an important component in many quantum algorithms. It is conjectured that it will be possible to generalise many of these results to the model introduced herein, however as the main focus here is on the one-way model this is left for future work. In later sections the following proposition will be required:

\begin{prop}
\label{cnotcircuits}
An $n$-qudit circuit consisting of only controlled-$X$ and controlled-$Z$ gates may be implemented with an unbounded fan-out circuit of $O(n^2)$ size and $O(1)$ depth.
\end{prop}
For brevity, the constructive proof of this proposition is relegated to Appendix~\ref{Aproof1} (which includes some further basic results which may be of minor significance to readers interested in qudit circuits). Although not a specific aim of the following investigations into the qudit one-way model, the results in the following section will provide a (constructive) proof of the following:
\begin{prop}\label{Pfanoutcliffdepth} Any Clifford operator on $n$ qudits may be implemented with an unbounded fan-out circuit of $O(n^4)$ size and $O(1)$ depth.
\end{prop}

\section{Qudit measurement patterns \label{dpatterns}}
The qudit one-way computer was first proposed by Zhou \emph{et al.} \cite{zhou2003quantum} in terms of measurements on cluster states and generalises the original binary logic model presented in Ref.~\cite{raussendorf2001one}. Although there has been much work investigating the computational power of the qubit one-way computational model and its relation to quantum circuits, for example see \cite{raussendorf2003measurement,danos2007measurement,Broadbent20092489,danos2009extended,browne2011computational}, these results have not been extended to multi-valued logic and a detailed understanding of the qudit one-way computational model remains to be developed. This is the topic of the remainder of this paper.

\subsection{Commands and patterns}
 The notation and terminology defined in the remainder of this paper is chosen to closely match that in common use for the qubits. Qudit \emph{measurement patterns} are now defined, using a similar notation to that introduced by Danos \emph{et al.} \cite{danos2007measurement} (for qubits). Such patterns will include the qudit cluster state model but are more general.
\newline
\indent
 The qudit one-way computer is defined here to be a model in which the allowed operations are the \emph{entangling commands}, \emph{Pauli corrections} and \emph{dependent} and \emph{independent measurements} which will be defined in-turn below. The set of preparable states is taken to be $\mathcal{S}=\{\ket{+_0}\}$. The \emph{entangling command} denoted $E_{ij}$, where $i$ and $j$ are the qudits on which it acts, is defined by
\begin{equation} E_{ij} := C^i_jZ, \end{equation}
which is simply the controlled-$Z$ operator. The \emph{Pauli corrections} are classically-controlled $X$ and $Z$ operators, specifically they are $X^{s_x}_i$ and $Z^{s_z}_i$ operators (acting on qudit $i$) where $s_x,s_z \in \mathbb{Z}(d)$ are classical dits calculated as the modulo $d$ sum of measurement outcomes (see below) and their additive inverses in $\mathbb{Z}(d)$ (the additive inverse of $a$ is $d-a$).
\newline
\indent
 The final type of commands are the measurement commands which output classical dits. For $\boldsymbol{\theta}$ and dits $s,t \in \mathbb{Z}(d)$, define the states
\begin{align} \ket{j_{(\boldsymbol{\theta},s,t)} } :=  v(\boldsymbol{\Lambda}(\boldsymbol{\theta},s,t) )^{\dagger}  \ket{j},
\end{align}
where the  $\boldsymbol{\Lambda}(\boldsymbol{\theta},s,t)$ is the vector with elements $\Lambda_j= \theta_{j+s} + tj\frac{2 \pi}{d} $. The dependent measurement command  $\tensor*[_t]{\left[M^{\boldsymbol{\theta } }_i \right]}{^s} $ is \emph{defined} to be a destructive measurement of the observable
\begin{equation} \hat{O}_i(\boldsymbol{\theta},s,t) = \sum_{j=0}^{d-1} j \ket{j_{(\boldsymbol{\theta},s,t)}} \bra{j_{(\boldsymbol{\theta},s,t)}}_i, \end{equation}
which hence measures qudit $i$ and outputs a single dit, denoted $s_i$. The term `destructive' should be taken to mean that the qudit is destroyed or discarded (and hence traced out) after the measurement. An \emph{independent measurement}, denoted $M^{\boldsymbol{\theta } }_i $, is defined by a measurement of $\hat{O}_i(\boldsymbol{\theta},0,0)$ and therefore does not depend on any classical dits.
\newline
\indent
 A measurement $\tensor*[_t]{\left[M^{\boldsymbol{\theta } }_i \right]}{^s} $ is equivalent to applying $v(\boldsymbol{\Lambda}(\boldsymbol{\theta},s,t) )$ followed by a computational basis measurement and furthermore, as up to a phase factor
\begin{equation}  \bra{j_{(\boldsymbol{\theta},s,t)} }  =\bra{j} v(\boldsymbol{\theta}) X^sZ^t,  \label{jv} \end{equation}
which may be confirmed with some simple algebra, then
\begin{equation}\tensor*[_{t+t'}]{\left[M^{\boldsymbol{\theta } }_i \right]}{^{s+s'}}= \tensor*[_t]{\left[M^{\boldsymbol{\theta } }_i \right]}{^s} X^{s'}_i Z^{t'}_i  = M^{\boldsymbol{\theta } }_i  X_i^{s+s'} Z_i^{t+t'} \label{deptoin} .\end{equation}
Hence it is simple to convert between dependent measurements and independent measurements preceded by Pauli corrections.

\subsection{Universal measurement patterns}
\begin{figure}
\includegraphics[width=6.0cm]{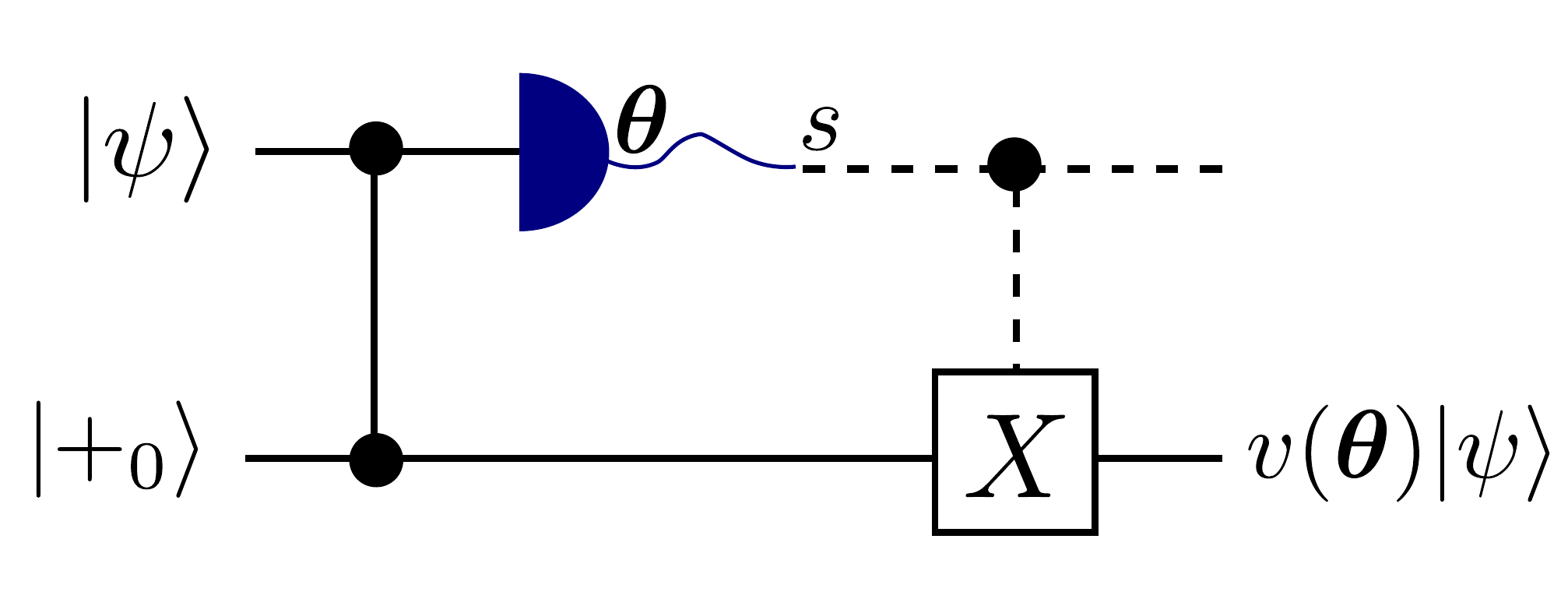}
\caption{\label{basicmp}A quantum-classical hybrid circuit representation of the measurement pattern for $\mathfrak{P}_{v(\boldsymbol{\theta})}$ as given in Eq.~(\ref{MpV}). The dotted line is the classical dit obtained as the output of the measurement.}
\end{figure}

A computation in the one way model is called a \emph{measurement pattern} (a particular pattern is specified by giving a quadruplet $\mathfrak{P}=(\mathcal{V},\mathcal{I},\mathcal{O},\mathfrak{p})$ where $\mathfrak{p}$ is a sequence of commands on $\mathcal{V}$). The definitions of the allowed commands in the model may appear rather technical and hence in order to illustrate how a measurement pattern implements a quantum computation, and to demonstrate the universality of the model, two examples of patterns are now given. An essentially trivial pattern is that for $CZ$. As $E_{ij}=CZ$ this can be implemented with the measurement-free pattern
 \begin{equation} \mathfrak{P}_{CZ} = (\{ 1,2\},\{1,2\},\{1,2\},  E_{1,2})\label{MpCZ}, \end{equation}
in which $\mathcal{V}=\mathcal{I}=\mathcal{O}$. A measurement pattern which implements $v(\boldsymbol{\theta})$ is
\begin{equation} \mathfrak{P}_{v(\boldsymbol{\theta})} = (\{ 1,2\},\{1\},\{2\}, X^{s_1}_2 M^{\boldsymbol{\theta}}_1 E_{1,2}). \label{MpV} \end{equation} 
That this indeed implements the required unitary can be shown with straightforward algebra and is essentially qudit teleportation as is made clear by Fig.~\ref{basicmp}. A derivation is included as Appendix~\ref{Asinglepattern} as an aid to further clarify the model.
Along with the definitions for composing computations, these two patterns demonstrate the universality of the model as any qudit unitary may be decomposed into a sequence of these operators. Note that the universality of the qudit one-way model has already been shown by Zhou \emph{et al.} \cite{zhou2003quantum} using the cluster state formalism. 
\subsection{Completely standard measurement patterns}
The presentation given above does not highlight the advantages over quantum circuits inherent in measurement patterns. These advantages can be illuminated by introducing a command rearranging process, termed \emph{standardisation} by Danos \emph{et al.} \cite{danos2007measurement} when introduced for qubits.

\subsubsection{Standardisation}
Composite measurement patterns can be rearranged so that they consist of an initial sequence of entangling commands, followed by dependent measurements and final Pauli corrections on the output qudits. This then links general measurement patterns to computation with cluster states, in which dependent measurements are performed on pre-prepared entangled states. Operations on distinct qudits commute and may be freely rearranged (with the exception that commands may not be rearranged so that an operation depends on a dit from a measurement yet to be performed). Hence, in combination with the rewrite rules
\begin{align}  E_{ij} \cdot X^s_i Z^t_i & \Rightarrow X^{s}_i Z^t_i Z^{s}_j \cdot E_{ij} ,\label{RewriteF} \\
 \tensor*[_t]{[M_{i}^{\boldsymbol{\theta}}]}{^s} \cdot  X^{s'}_i Z^{t'}_i & \Rightarrow  \tensor*[_{t+t'}]{[M_{i}^{\boldsymbol{\theta}}]}{^{s+s'}}  \label{RewriteE}, \end{align}
 the commands in a pattern may be reordered as described above. It is important to note that these rewrite rules do \emph{not} change the output of the computation as they hold as equalities by Eq.~(\ref{DcliffCZ}) and Eq.~(\ref{deptoin}) respectively.

\subsubsection{Pauli simplification}
In the case of patterns including Clifford operators, a further stage of pattern rewriting can be implemented called \emph{Pauli simplification}. The patterns $\mathfrak{P}_{CZ}$, $\mathfrak{P}_{F}=\mathfrak{P}_{v(\boldsymbol{0})}$ and $\mathfrak{P}_{FP} = \mathfrak{P}_{v(\boldsymbol{p})}$ where $p_j = \pi j ( j+\delta_d)/d$ ($\delta_d=0,1$ for even and odd $d$ respectively) generate the Clifford group (see Eq.~(\ref{Cliffgen})). For the two measurements required to implement these patterns introduce the rewrite rules
\begin{align} \label{Paulirewrite} \tensor*[_t]{\left[M^{\boldsymbol{0} }_i \right]}{^s} & \Rightarrow \tensor*[_t]{\left[M^{\boldsymbol{0} }_i \right]}{}, \\  \tensor*[_t]{\left[M^{\boldsymbol{p} }_i \right]}{^s} & \Rightarrow \tensor*[_{s+t}]{\left[M^{\boldsymbol{p} }_i \right]}{}. \label{Pauli2} \end{align}
Again, these do not change the output of the computation as these hold as equalities for the following reasons: The operator that is measured is fixed by the vector $\boldsymbol{\Lambda}(\boldsymbol{\theta},s,t)$ with $\Lambda_j= \theta_{j+s} + tj\frac{2 \pi}{d} $. When $\boldsymbol{\theta}=(0,0,...0)$ this is clearly independent of $s$ and hence this dependence may be dropped. Using Eq.~(\ref{jv}) and the conjugation rules for the Fourier and phase gates given in Eq.~(\ref{DcliffF}) and (\ref{DcliffP}) it is simple to confirm that
$ \bra{j_{(\boldsymbol{p},s,t)} } =  \bra{j} FP X^sZ^t =   \bra{j}  Z ^{-s} FPZ^{s+t} = \bra{j}  FPZ^{s+t} = \bra{j_{(\boldsymbol{p},0,s+t)} }$ with each equality holding up to a phase. Hence $\hat{O}_i(\boldsymbol{p},s,t) = \hat{O}_i(\boldsymbol{p},0,s+t)$ confirming that (\ref{Pauli2}) holds as an equality as $\tensor*[_t]{\left[M^{\boldsymbol{p} }_i \right]}{^s}$ is defined as the measurement of $\hat{O}_i(\boldsymbol{p},s,t) $.

\subsubsection{Signal shifting}
The final rewrite rule to be introduced is termed \emph{signal shifting} and removes \emph{all} Pauli-$Z$ dependencies in \emph{all} the measurements via the replacement
\begin{equation} s_i, \tensor*[_t]{[M_{i}^{\boldsymbol{\theta}}]}{^s}   \Rightarrow s_i-t, [M_{i}^{\boldsymbol{\theta}}]^s .\end{equation}
This denotes that the removal of the $Z$-type dependency on dit $t$ from a measurement of qudit $i$ is accompanied by replacing any dependencies on $s_i$ with dependencies on $s_i -t$ (this clearly introduces classical computation). Again this may be confirmed to leave the output of the computation unchanged by observing that $ \bra{j_{(\boldsymbol{\theta},s,t)} } = \bra{j} v(\boldsymbol{\theta}) X^sZ^t = \bra{j}  X ^{-t} v(\boldsymbol{\theta})X^s =\bra{j+t} v(\boldsymbol{\theta})X^s =  \bra{(j+t)_{(\boldsymbol{\theta},s,0)} }$ with each equality only up to a phase. Hence \begin{equation} \hat{O}_i(\boldsymbol{\theta},s,t) = \sum_{j=0}^{d-1} (j-t) \ket{j_{(\boldsymbol{\theta},s,0)}} \bra{j_{(\boldsymbol{\theta},s,0)}}_i, \end{equation}
which is clearly exactly equivalent to a measurement of $\hat{O}_i(\boldsymbol{\theta},s,0)$ along with the classical post-processing of subtracting $t$ modulo $d$ from the outcome.
\newline
\indent
We will call the composite process of standardisation, Pauli simplification and then signal shifting \emph{complete standardisation} and a pattern on which this has been applied will be called \emph{completely standard}. An example of completely standardising a measurement pattern is given in Appendix~\ref{Aexstand}. Complete standardisation never increases the (quantum) size or depth of a pattern but in general it adds the requirement for simple classical processing in the form of addition modulo $d$. A proof of this is not included as it may be shown in a similar fashion to the equivalent result for qubits derived in Ref. \cite{danos2007measurement} . It is easily confirmed that a completely standard Clifford measurement pattern will have no dependent measurements and hence all of the measurements may be performed simultaneously.

\begin{definition}
 The entanglement depth is the minimum depth of the entanglement commands in a standardised pattern. 
\end{definition}
We take it to be the \emph{minimum} depth as by arranging the entangling commands in a particularly inconvenient order the depth can obviously be increased, but as they can be freely commuted it is more useful to know what the minimum depth can be by a judicious rearrangement of these commands (consider a `cascade' of $CZ$ gates which may be arranged for the depth to be either two or the same as the number of gates).
The entanglement stage of a pattern may be represented conveniently as a unique graph, in which the nodes are the qudits and the number of edges between nodes represent the number of entangling commands acting on each qudit pair (and hence the number of edges between two nodes should be in $\mathbb{Z}(d)$ as $(CZ)^d = \mathbb{I}$).
\begin{lemma} (\cite{Broadbent20092489} Lemma 3.1) Let $G$ be the entanglement graph of a standardised pattern $\mathfrak{P}$ and let  $\Delta(G)$ be maximum degree of $G$. The entanglement depth of $\mathfrak{P}$ is either $\Delta(G)$ or $\Delta(G)+1$.
\end{lemma}
The lemma of Ref.~\cite{Broadbent20092489} is presented in the context of qubit measurement patterns but as it only relates to the properties of the entanglement graph, it is easily confirmed that it also applies here.

\subsection{Converting between measurement patterns and quantum circuits}
We now present mappings in both directions between qudit quantum circuits and measurement patterns (see Ref. \cite{Broadbent20092489} for similar work with qubits). This will then be used to provide depth-preserving mappings between measurement patterns and unbounded fan-out circuits, generalising a result of Browne \emph{et al.} \cite{browne2011computational}  to multi-valued logic.

\subsubsection{Measurement patterns simulating quantum circuits}
\begin{definition} \label{patternfromcircuit} The standard measurement pattern simulation of a quantum circuit is obtained by
\begin{enumerate}
\item Rewriting the circuit as the composition of single-gate circuits $\mathfrak{C}_{CZ}$ and $\mathfrak{C}_{v(\boldsymbol{\theta})}$
\item Replacing each basic circuit in the decomposition with the equivalent basic measurement patterns $\mathfrak{P}_{CZ}$ and  $\mathfrak{P}_{v(\boldsymbol{\theta})}$
\item Completely standardising the resultant measurement pattern.
\end{enumerate}
\end{definition}
It is noted that this procedure introduces additional ancillary qudits.

\begin{lemma} \label{Pctom} Any standard quantum circuit $\mathfrak{C}$ may be implemented with a measurement pattern $\mathfrak{P}$ with depth $O(\text{depth} (\mathfrak{C}))$ and $O(\text{size}(\mathfrak{C}))$.
\end{lemma}
Consider the standard measurement pattern implementation of the circuit. Each basic measurement pattern replacing each basic gate is at most a small constant increase in size and depth.
\newline
\indent
This method of converting a quantum circuit into a measurement pattern is not in general optimal in terms of the depth of the pattern and will not always give constant depth Clifford patterns. Consider for example any circuit consisting of only $CZ$ gates, in which case the measurement pattern will include no measurements and have an identical depth to the circuit. Hence an alternative procedure is now given.
\begin{definition}
\label{clusterdef}
The cluster-state measurement pattern simulation of a quantum circuit is found using an identical procedure to the standard measurement pattern except that before conversion to a measurement pattern, four $\mathfrak{C}_F$ basic circuits are inserted between any $\mathfrak{C}_{CZ}$ gates that act consecutively on the same qudit.
\end{definition}
This has no effect on the unitary implemented by the circuit (and hence the resultant measurement pattern) as $F^4=\mathbb{I}$ and will increase the depth and size of the circuit (and pattern) by less than a factor of four. However, it can easily be confirmed that now the entanglement graph of the pattern has nodes of at most degree three.

\begin{prop} \label{Pcliffdepth} Any Clifford operator on $n$ qudits may be implemented with an $O(n^2)$ size and constant depth measurement pattern.
\end{prop}
Any Clifford gate on $n$ qudits may be decomposed into a circuit that requires no ancillas and consists of $O(n^2)$ $F$, $P$ and $CZ$ gates using the algorithm of Farinholt \cite{farinholt2014ideal}. Consider the cluster-state measurement pattern simulation of this circuit. This pattern still has a size of $O(n^2)$. Such a pattern has a constant depth for the entanglement operations (at most $4$). As the pattern is completely standard and the measurement angles are only those for implementing $F$ and $FP$ all the measurements are independent and hence may all be implemented simultaneously requiring unit depth. The corrections all apply to different qudits in the output and hence may be applied in a depth of 2. 
\begin{lemma} \label{fanoutmp} The qudit fan-out operator on $n$ qudits can be implemented with a constant depth and $O(n)$ size measurement pattern.
\end{lemma}
The qudit fan-out gate is Clifford as can be seen from its decomposition into controlled-$X$ gates in Fig.~\ref{quditfanout}. Hence by proposition~\ref{Pcliffdepth} this may be implemented in constant depth. The  $O(n)$ size scaling is obvious from the proof of the above proposition.
\begin{lemma} 
\label{fantopat}
Any unbounded fan-out circuit $\mathfrak{F}$ may be implemented with a measurement pattern $\mathfrak{P}$ of depth $O(\text{depth} (\mathfrak{F}))$ and size $O(\text{size}(\mathfrak{F}))$. 
\end{lemma}
This follows from lemmas~\ref{Pctom} and~\ref{fanoutmp}.

\subsubsection{Circuit simulations of measurements patterns}
\begin{definition} \label{cohpatt} The coherent circuit simulation of the measurement pattern $\mathfrak{P}=(\mathcal{V},\mathcal{I},\mathcal{O},\mathfrak{p})$ is the circuit $\mathfrak{C}=(\mathcal{V},\mathcal{I},\mathcal{O},\mathfrak{c}(\mathfrak{p}))$ where $\mathfrak{c}(\mathfrak{p})$ consists of an initial layer of $F$ gates on all qudits in $\mathcal{V} \setminus \mathcal{I}$ followed by the commands of $\mathfrak{p}$ in order using the replacements
\begin{enumerate}
\item $E_{i,j}$ $\Rightarrow C^i_jZ$,
\item $M^{\boldsymbol{\theta } }_i$  $\Rightarrow v(\boldsymbol{\theta})_i,$ 
\item $X^{\sum_{i \in \mathbb{S} } \pm_i s_i}_j \Rightarrow \prod_{i \in \mathbb{S}} C^{i}_jX^{\pm_i 1}$,
\item $Z^{\sum_{i \in \mathbb{S} } \pm_i t_i}_j \Rightarrow \prod_{i \in \mathbb{S}} C^{i}_jZ^{\pm_i 1}$.
\end{enumerate}
\end{definition}
This may be used to turn any measurement pattern into a quantum circuit by decomposing any dependent measurements into Pauli corrections and independent measurements. This method for the coherent implementation of a measurement pattern explicitly highlights the intrinsic role of classical computation in the one-way model: In steps 3 and 4 local gates controlled by \emph{classically calculated} dit sums (modulo $d$) are replaced by a sequence of two-qudit gates in which these sums are calculated using unitary \emph{quantum gates}. Hence, the power of the one-way model is in using classical computation instead of quantum computation when the quantum element is superfluous.
\begin{prop} \label{patttocirc} Any measurement pattern $\mathfrak{P}$ may be implemented with an unbounded fan-out circuit with a depth of $O(\mathrm{depth} (\mathfrak{P}))$ and a size of $O(\mathrm{size}(\mathfrak{P})^2\mathrm{depth} (\mathfrak{P}))$.
\end{prop}
Consider a completely standard pattern $\mathfrak{P}$ \cite{noteMBQC_D}. The command sequence of $\mathfrak{P}$ consists of three sequential stages: (i) entanglement operations; (ii) Pauli-$X$ dependent measurements; (iii) Pauli-$X$ and Pauli-$Z$ corrections on the output qudits. Consider the coherent circuit implementation of $\mathfrak{P}$ as given by definition~\ref{cohpatt}. The preliminary stage of this circuit consists of $F$ gates on the qudits in $\mathcal{V} \setminus \mathcal{I}$. This may be implemented by an unbounded fan-out circuit with unit depth and a size no greater than $\text{size}(\mathfrak{P})$. Consider stage (i): This consists of $CZ$ gates which have the same size and depth in the measurement pattern and with an unbounded fan-out circuit. Consider stage (ii): This circuit subsection consists of no more than $\text{depth}(\mathfrak{P})$ layers which are each formed from controlled-$X$ gates act on at most $\text{size}(\mathfrak{P})$ qudits (from the Pauli-$X$ corrections), followed by some local gates which all act on distinct qudits (from the independent measurements). Any circuit acting on $\text{size}(\mathfrak{P})$ qudits and consisting of only controlled-$X$ gates may be implemented with an unbounded fan-out circuit of size $O(\text{size}(\mathfrak{P})^2)$ and depth of $O(1)$ by proposition~\ref{cnotcircuits}. The local gates may be implemented with unit depth. As there are at most $\text{depth}(\mathfrak{P})$ such layers this results in a total size for this stage of the circuit of $O(\text{size}(\mathfrak{P})^2\text{depth}(\mathfrak{P}))$ and a depth of $O(\text{depth}(\mathfrak{P}))$. Consider stage (iii): This is a sequence of controlled-$X$ and Pauli-$Z$ gates on at most $\text{size}(\mathfrak{P})$ qudits. By proposition~\ref{cnotcircuits} this may also be implemented by an unbounded fan-out circuit of size $O(\text{size}(\mathfrak{P})^2)$ and depth of $O(1)$. Hence, the unbounded fan-out circuit simulation of $\mathfrak{P}$ has a size of $O(\text{size}(\mathfrak{P})^2\text{depth}(\mathfrak{P}))$ and a depth of $O(\text{depth}(\mathfrak{P}))$.
\newline
\indent
Finally it is noted that combining this result with proposition~\ref{Pcliffdepth} proves the earlier claim about the size and depth complexity of unbounded fan-out Clifford circuits given in proposition~\ref{Pfanoutcliffdepth}.

\section{Complexity classes}
A simple way in which to summarise these results is in terms of complexity classes. For quantum circuits with \emph{qubits}, the complexity class $\textsc{QNC}^k$ of operators (or alternatively decision problems) that may be computed by poly-logarithmic depth ($O(\log^k n)$) standard quantum circuits was first introduced by Moore and Nilsson as the quantum analog of the equivalent classical circuit class $\textsc{NC}^k$ \cite{moore2001parallel}. Equivalent complexity classes for qubit unbounded fan-out circuits, denoted $\textsc{QNC}_f^k$ \cite{hoyer2003quantum} and measurement patterns, denoted $\textsc{QMNC}^k$ \cite{browne2011computational} have also been defined, and these classes are now generalised to qudit circuits.
 
 \begin{definition}  The complexity classes $\textsc{QNC}^k_d$, $\textsc{QNC}^k_{f,d}$ and $\textsc{QMNC}^k_d$ contain operators computed exactly by uniform families of qudit standard quantum circuits, unbounded fan-out circuits and measurement patterns respectively which have input size $n$, depth $O(\log^k n)$ and polynomial size.
\end{definition}
Proposition~\ref{logdepthfanout} implies the complexity class inclusion
\begin{equation}  \textsc{QNC}^0_d \subset  \textsc{QNC}_{f,d}^0 \subseteq  \textsc{QNC}_{d}^{1},  \end{equation}
which summarises the difference in depth complexity between standard quantum circuits and unbounded fan-out circuits. For all $k$, $ \textsc{QNC}^k_d \subseteq \textsc{QNC}_{f,d}^k \subseteq \textsc{QNC}_{d}^{k+1}$ but it has not been shown whether for $k \neq 0$ any of these inclusions are strict (this is also the case for qubits \cite{browne2011computational}). The relationship between unbounded fan-out circuits and measurement patterns that has been provided in lemma~\ref{fantopat} and proposition~\ref{patttocirc} can be summarised by
\begin{equation} \textsc{QNC}_{f,d}^k  =  \textsc{QMNC}^k_d, \end{equation}
which is a generalisation of a theorem of Browne \emph{et al.} \cite{browne2011computational} to higher dimensions. Finally, it is noted that no quantum computational model in which the non-input states are initialised in Pauli eigenstates and the only allowed operations are Clifford unitaries, single-qudit measurements and Pauli corrections controlled by the sum modulo $d$ of sets of these measurement outcomes, can have a lower depth complexity than measurement patterns. This generalises a result of Browne \emph{et al.} \cite{browne2011computational} for qubits and is shown in Appendix~\ref{Opt} (the result given is more general than this and applies to multi-qudit measurements). Interestingly, this shows that the qudit ancilla-driven model recently presented by this author and Kendon \cite{Proctor2015ancilla} has the same depth complexity as measurement patterns. This is because in Ref.~\cite{Proctor2015ancilla} a depth preserving map is provided from qudit measurement patterns to the model therein.
\section{Conclusions}
In this paper it has been shown that the qudit one-way model is powerful for low-depth quantum computation. In order to illuminate the relationship between this model and qudit quantum circuits an `unbounded fan-out' model was first introduced, in which qudits may be quantum-copied into any number of ancillas in a single-time step, and it was shown that this model may implement interesting gate sequences in low depth. This model generalises the well-studied qubit unbounded fan-out model \cite{hoyer2005quantum,hoyer2003quantum,takahashi2013collapse,takahashi2010quantum,moore2001parallel} to higher dimensions and interesting future work would be to fully investigate its parallel computational power.
\newline
\indent
Qudit measurement patterns were then introduced, which adapt the cluster-state qudit model of Zhou \emph{et al.} \cite{zhou2003quantum} to a more flexible setting well-suited to a comparison with the gate model. Depth reduction `standardisation' protocols were then developed, in a similar vein to the qubit work of Danos \emph{et al.} \cite{danos2007measurement}, and in doing so a simple procedure for mapping between quantum circuits and measurement patterns was then provided. Using this it was shown that the depth complexity of the qudit one-way model is exactly equivalent to the qudit unbounded fan-out model, confirming and making precise the parallelism inherent in the qudit one-way model.
\newline
\indent
The procedures introduced herein for turning a measurement pattern into a completely unitary quantum circuit explicitly highlight that the power of the one-way model is in replacing parts of the quantum computation with the exactly equivalent \emph{classical} computation. This essentially uses the well-known result that quantum controlled gates followed by a computational basis measurement of the control subsystem may simply be replaced by an earlier measurement and classically controlled gates \cite{nielsen2010quantum}. It is a systematic use of this principle along with ancillary qudits and a judicious re-ordering procedure that provides the one-way model with this parallelism. Hence, if all resources (i.e., quantum and classical) are treated on an equal footing this improved parallelism in comparison to unitary circuits disappears. However, from a practical perspective it is clear that to at least some degree, quantum and classical resources should be counted on a different footing.
\newline
\indent
Finally, it noted that further interesting work would be to investigate how the full range of highly-developed concepts in qubit measurement-based quantum computation may be generalised to higher dimensions, for example information flow notions \cite{browne2007generalized,duncan2010rewriting} or the inter-play between universal quantum and classic computational resources \cite{anders2009computational}.

\section*{Acknowledgments}
The author would like to thank Viv Kendon for a careful reading of the manuscript and Dan Browne for interesting discussions related to this work and comments on the manuscript. The author was funded by a University of Leeds Research Scholarship.

\appendix
\section{ \label{Aproof1}}
This appendix contains a proof of proposition~\ref{cnotcircuits}. In order to simplify the presentation of the proof it will be useful to first introduce some further $n$-qudit gates.
\subsection{Generalised fan-out and modulo $d$ gates}
A gate which is closely related to the qudit fan-out gate and which will be called the \emph{modulo-}$d$ gate may be defined by
\begin{equation*} \textsc{mod} \ket{x} \ket{y_1,...,y_{n}} :=\left| x+ y_1+...+y_n \right\rangle \ket{y_1,...,y_{n}} , \end{equation*} 
This is the natural generalisation of the qubit parity gate to higher dimensions. It is easily confirmed via the conjugation relations for $F$ in Eq.~(\ref{DcliffF}) that
\begin{equation} F^{\otimes^{n+1}} \textsc{fanout}(F^{\dagger})^{\otimes^{n+1}}  = \textsc{mod}^{-1}. \end{equation}
which has the same form as the well-known relationship between fan-out and parity for qubits first noted by Moore \cite{moore1999quantum}. The qubit fan-out and parity gates are self-inverse, however for general $d$ they have order $d$, i.e., $\textsc{fanout}^d=\textsc{mod}^d=\mathbb{I}$. Due to this, a useful extension of these gates (that is trivial for $d=2$)  is the $\boldsymbol{v}=(v_1,v_2,...,v_n) \in \mathbb{Z}(d)^{n}$ parameterised \emph{generalised fan-out} gate:
\begin{equation*} \textsc{fanout}(\boldsymbol{v}) \ket{x} \ket{y_1,...,y_{n}}: = \ket{x}   \left| y_1+v_1x,...,y_n+ v_{n}x \right\rangle,\end{equation*} 
and the \emph{generalised modulo-}$d$ gate:
\begin{equation*} \textsc{mod}(\boldsymbol{v}) \ket{x} \ket{y_1,...,y_{n}} :=\left| x+ v_1y_1+...+v_ny_n  \right\rangle \ket{y_1,...,y_{n}} . \end{equation*} 
It is easy to confirm the relation:
\begin{equation} F^{\otimes^{n+1}} \textsc{fanout}(\boldsymbol{v}) (F^{\dagger})^{\otimes^{n+1}}  = \textsc{mod}(-\boldsymbol{v}). \label{fanmod} \end{equation}
\begin{lemma} \label{modgates} Any $n$-qudit generalised fan-out or modulo-$d$ gate may be implemented in a depth of $O(1)$ and size of $O(n)$ with an unbounded fan-out circuit.
\end{lemma}
It is only necessary to show how to implement any generalised fan-out gate as then a generalised modulo-$d$ gate may be implemented by the relation in Eq.~(\ref{fanmod}). A generalised fan-out may be implemented with a standard fan-out and controlled-$X$ gates as follows. Fan-out the control qudit into $n-1$ copies. Use these to implement $C_j(X^{v_j})$ gates in parallel (each gate is on a distinct qudit) using controlled-$X$ gates. Apply fan-out $d-1$ times to inverse the fan-out of the control qudit.  Each stage has a depth which at worst scales with $d$.
\subsection{Constant depth controlled Pauli circuits }
\begin{figure*}
\includegraphics[width=18cm]{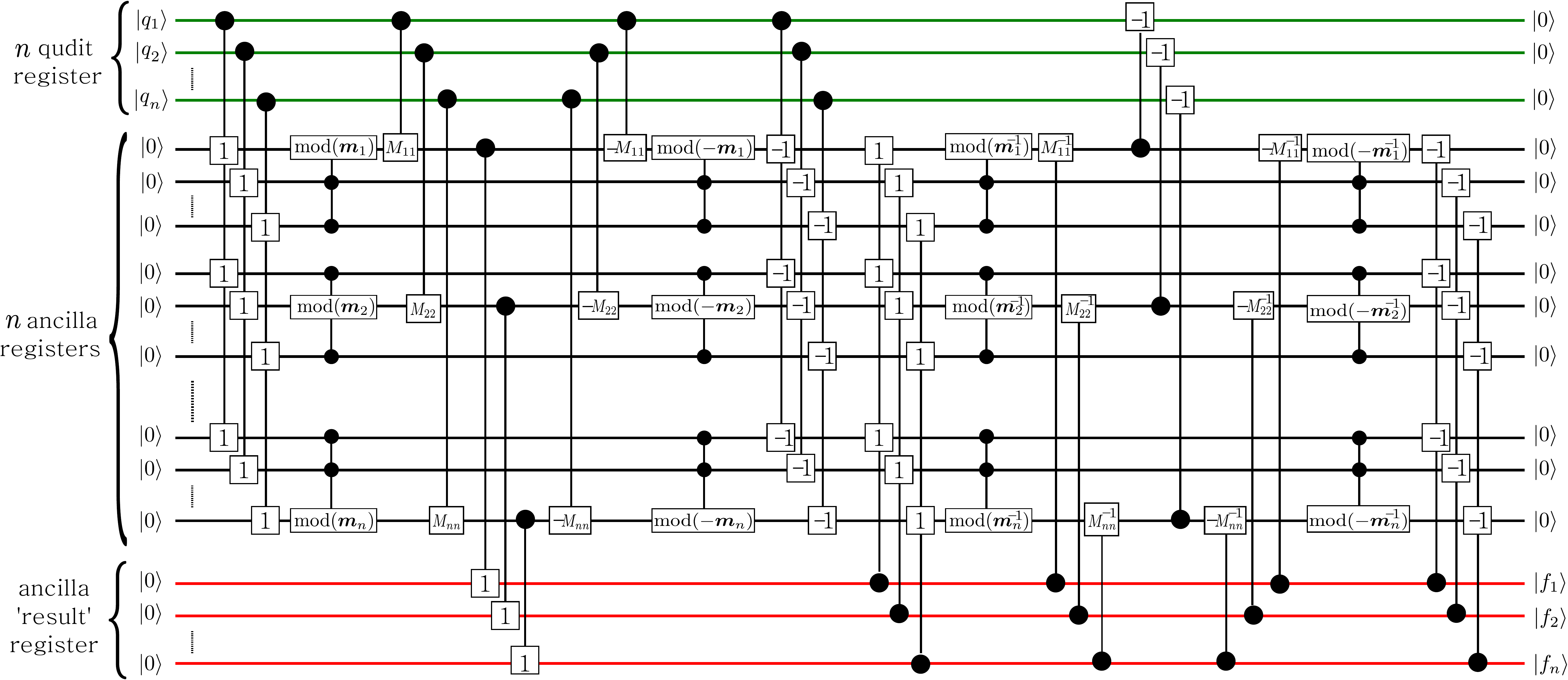}
\caption{\label{CXconstdepth}(color online) An unbounded fan-out circuit of $O(n^2)$ size and $O(1)$ depth that implements \emph{any} circuit on $n$ qudits consisting of only controlled-$X$ gates. The parameters in this circuit are found as described in the proof of Proposition~\ref{cnotcircuits}. This does not include the final \textsc{swap} stage (see the proof of Proposition~\ref{cnotcircuits}) and hence the output of this circuit is found in the `result' ancillary register with all other qudits in the state $\ket{0}$.}
\end{figure*}

Proposition~\ref{cnotcircuits} is now proven, which states that an $n$-qudit circuit consisting of only controlled-$X$ and controlled-$Z$ gates may be implemented with an unbounded fan-out circuit of $O(n^2)$ size and $O(1)$ depth. The proof is given in stages, the first of which involves a reordering of the gates in the circuit.
\newline
\indent
\emph{1. Rearranging the circuit:}
To begin, the circuit is rearranged so that it consists of a sequence of controlled-$Z$ and ordinary $Z$ gates, followed by a sequence of controlled-$X$ gates. Clearly gates on distinct qudits commute and hence it is only necessary to provide a rule for commuting gates which act on at least one qudit in common. As $CZ$ is symmetric there are only three cases to consider: $C^i_j (Z) C^i_j (X)$, $C^i_j(Z) C^i_k(X)$ and $C^i_j (Z) C^k_j (X)$. It is easily confirmed using the Weyl commutation relation (and the relation $C^i_j(\omega \mathbb{I}) = Z_i$) that
\begin{align}
C^i_j (Z) C^i_j (X)& =  C^i_j (X) C^i_j (Z) Z_i ,\\
C^i_j (Z) C^i_k (X)& =  C^i_k (X) C^i_j (Z),  \\
C^i_j (Z) C^k_j (X)& =  C^k_j (X) C^i_j (Z) C^i_k (Z) .
 \end{align}
The first of these relations creates a local $Z$ operator when used to reorder the gates, however these local gates may easily be commuted through further controlled-$X$ gates with the relations 
\begin{align} Z_i C^i_j(X) &=C^i_j(X)Z_i,\\
Z_j C^i_j(X) &=C^i_j(X) Z_iZ_j .\end{align} Hence it is possible to rearrange the circuit into local $Z$ and controlled-$Z$ gates followed by controlled-$X$ gates.
\newline
\indent
\emph{2. Implementing the controlled-$Z$ sub-circuit:} This part of the circuit is diagonal. By proposition~\ref{diagprop}, this may be implemented with a constant depth and $O(n^2)$ size unbounded fan-out circuit.
\newline
\indent
\emph{3. A representation of the controlled-$X$ sub-circuit:} Controlled-$X$ gates map computational basis states to computational basis states. More specifically, they map
\begin{equation} \ket{q_1,...,q_n} \to \ket{f_1,...,f_n}\end{equation}
 where for some $M_{jk} \in \mathbb{Z}(d)$, $j,k=1,..,n$, each $f_j$ is given by
\begin{equation} f_j= M_{1j} q_1 +M_{2j} q_2 +  ...+M_{nj} q_n\hspace{0.2cm}\text{mod} \hspace{0.1cm}d. \end{equation}
Hence the circuit may be represented by an $n \times n$ matrix $M$ with entries in $\mathbb{Z}(d)$. For a given circuit this matrix $M$ may be found in the following way. Start with the vector $(q_1,...,q_n)$ and consider each layer of gates in turn. For each controlled-$X$ update the vector by adding the current value of the control qudit in this vector to that of the target qudit. The final vector will obviously be $(f_1,...,f_n)$ and will provide the elements of $M$. The elements of $M$ may therefore be found with simple algebra.
\newline
\indent
\emph{4: Implementing the Pauli-$X$ sub-circuit:}
For any given $M$, which is found as above, it is now shown how to implement the associated controlled-$X$ circuit in constant depth and quadratic size with an unbounded fan-out circuit. The first steps is to make $n$ copies of the qudit register using $n$ fan-out gates (in parallel) and using $n^2$ ancilla qudits (initialised to $\ket{0}$). In the $k\textsuperscript{th}$ ancillary register the $k\textsuperscript{th}$ qudit is mapped from $q_k \to f_k$ using a generalised modulo$-d$ gate (the $k\textsuperscript{th}$ qudit in that register is the target and the remaining $n-1$ qudits are the control qudits. The gate is $\textsc{mod}(\boldsymbol{m}_{k})$ where $\boldsymbol{m}_{k} \in \mathbb{Z}(d)^{n-1}$ is the $k\textsuperscript{th}$ row of $M$ with the $M_{kk}$ element removed.) and a $(CX)^{M_{kk}}$ gate with the control the $k\textsuperscript{th}$ qudit in the original register. These gates may be implemented on each ancillary register in parallel and by lemma~\ref{modgates}, the generalised modulo-$d$ gate may be implemented via fan-out in constant depth and linear size. The value of each $f_k$ may be written into a further `result' ancillary register in a depth of 1 (using $n$ controlled-$X$ gates). The next step is to disentangle the $n$ ancillary registers from the original register and the `result' register by uncomputing $f_j$. This is achieved by applying the entire circuit (except the copying into the `result' register) backwards. This leaves $n^2$ clean ancillary registers with the original and `result' registers in the state $\sum \ket{q_1...q_n}\ket{f_1...f_n}$. 
\newline
\indent
The penultimate stage is to clean the original register (transform it into $\ket{0,0...0}$). To do this, $\ket{q_1...q_n}$ is calculated from $\ket{f_1...f_n}$ using the above method again (i.e., via the $n^2$ ancillary registers) but with the following changes: The roles of the original and result registers are reversed and $M$ is replaced with $M^{-1}$ (which may be easily found via the reverse of the procedure for finding $M$). This then computes $q_k$ on the $k\textsuperscript{th}$ qudit of the $k\textsuperscript{th}$ ancillary register. A controlled $X^{-1}$ gate on the $k\textsuperscript{th}$ qudit of the original register with the control qudit the $k\textsuperscript{th}$ qudit of the $k\textsuperscript{th}$ ancillary register maps the target to $\ket{q_k-q_k}=\ket{0}$. As above, the inverse computation is implemented to unentangle the $n$ ancillary registers leaving the original and result registers in the state $\ket{0,0...0}\ket{f_1,f_2,...,f_n}$. The circuit is completed by swapping the original and `result' registers which may be implemented by $n$ $\textsc{swap}$ gates in parallel, where $\textsc{swap}$ is defined by $\textsc{swap}\ket{n}\ket{m} := \ket{m}\ket{n}$. This may be implemented by three controlled Pauli gates \cite{garcia2013swap} and hence swapping the registers requires constant depth and linear size. The controlled-$X$ stage is therefore implementable by a unbounded fan-out circuit with a depth of $O(1)$ and size of $O(n^2)$. Combining the results in each of these stages concludes the proof. In order to clarify the controlled-$X$ sub-circuit a circuit diagram demonstrating the method is given in Fig.~\ref{CXconstdepth}.

\section{ \label{Asinglepattern}}
In this appendix it is confirmed that the measurement pattern
\begin{equation} \mathfrak{P}_{v(\boldsymbol{\theta})} = (\{ 1,2\},\{1\},\{2\}, X^{s_1}_2 M^{\boldsymbol{\theta}}_1 E_{1,2}), \end{equation} 
does indeed implement the unitary transformation $v(\boldsymbol{\theta})$ on an arbitrary input. This therefore requires showing that
\begin{equation} \ket{\Psi_j} \equiv X^{s_1}_2 M^{\boldsymbol{\theta}}_1 E_{1,2} \ket{j}_1 \ket{+_0}_2 = v(\boldsymbol{\theta})\ket{j}_2, \label{pat1} \end{equation}
for all $j \in \mathbb{Z}(d)$, as by linearity it is only necessary to show that the procedure implements the required map on every computational basis state. Eq.~(\ref{pat1}) is now derived. From the definition of the entangling command and the action of $Z$ on the conjugate basis it follows that
\begin{align*} \ket{\Psi_j}
 & = X^{s_1}_2  M^{\boldsymbol{\theta}}_1CZ \ket{j}_1 \ket{+_0}_2, \\
& = X^{s_1}_2  M^{\boldsymbol{\theta}}_1 \ket{j}_1 \ket{+_j}_2 .
\end{align*}
The measurement command is then equivalent to projecting qudit 1 onto the state $v(\boldsymbol{\theta})^{\dagger} \ket{s_1}$ (where $s_1 \in \mathbb{Z}(d)$ is the arbitrary measurement outcome) and renormalising, and hence
\begin{align*}
\ket{\Psi_j} & = X^{s_1}_2   \bra{s_1} v(\boldsymbol{\theta}) \ket{j}|\bra{s_1} v(\boldsymbol{\theta}) \ket{j}|^{-1} \ket{+_j}_2 \\
& = X^{s_1}_2  e^{i \theta_j}  \expect{s_1, +_j} | \expect{s_1, +_j }|^{-1} \ket{+_j}_2 
\end{align*}
Using the overlap of conjugate and computational basis states given in Eq.~(\ref{mub}) and the action of $X$ on the conjugate basis it is then clear that
\begin{align*}
\ket{\Psi_j} & = X^{s_1}_2  e^{i \theta_j}\omega^{s_1 j} \ket{+_j}_2 \\
& = X^{s_1}_2  X^{-s_1}_2   v(\boldsymbol{\theta}) \ket{j}_2 \\
& = v(\boldsymbol{\theta}) \ket{j}_2,
\end{align*}
which confirms Eq.~(\ref{pat1}).
\section{\label{Aexstand}}
In this appendix an example of composing measurement patterns and applying complete standardisation is given. Consider the single qudit unitary $v(\boldsymbol{\psi})v(\boldsymbol{\phi})v(\boldsymbol{\theta})$ which is implemented by the measurement pattern $\mathfrak{P}_{v(\boldsymbol{\psi})v(\boldsymbol{\phi})v(\boldsymbol{\theta})} = \mathfrak{P}_{v(\boldsymbol{\psi})} \circ \mathfrak{P}_{v(\boldsymbol{\phi)}} \circ \mathfrak{p}_{v(\boldsymbol{\theta})}$ where the measurement pattern $\mathfrak{P}_{v(\boldsymbol{\theta})}$ is given in Eq.~(\ref{MpV}). The composite measurement pattern is given (via the definition of composition) by
\[
 \mathfrak{P}_{v(\boldsymbol{\psi})v(\boldsymbol{\phi})v(\boldsymbol{\theta})} =  (\{ 1,2,3,4\},\{1\},\{4\},  \mathfrak{p}_{v(\boldsymbol{\psi})} \mathfrak{p}_{v(\boldsymbol{\phi})}\mathfrak{p}_{v(\boldsymbol{\theta})})
\]
where $\mathfrak{p} \equiv  \mathfrak{p}_{v(\boldsymbol{\psi})} \mathfrak{p}_{v(\boldsymbol{\phi})}\mathfrak{p}_{v(\boldsymbol{\theta})}$ is given by
\[\mathfrak{p} =  X^{s_3}_4 M^{\boldsymbol{\theta}}_3 E_{3,4}X^{s_2}_3 M^{\boldsymbol{\theta}}_2 E_{2,3} X^{s_1}_2 M^{\boldsymbol{\theta}}_1 E_{1,2}. \]
First apply standardisation to this sequence of commands. This procedure gives
\begin{align*}\mathfrak{p} & =  X^{s_3}_4 M^{\boldsymbol{\psi}}_3 E_{3,4}X^{s_2}_3 M^{\boldsymbol{\phi}}_2 E_{2,3} X^{s_1}_2 M^{\boldsymbol{\theta}}_1 E_{1,2}\\ 
 & \Rightarrow  X^{s_3}_4 M^{\boldsymbol{\psi}}_3 X^{s_2}_3Z^{s_2}_4 E_{3,4} M^{\boldsymbol{\phi}}_2  X^{s_1}_2  Z^{s_1}_3 E_{2,3} M^{\boldsymbol{\theta}}_1 E_{1,2}\\
  & \Rightarrow  X^{s_3}_4Z^{s_2}_4 [M^{\boldsymbol{\psi}}_3]^{s_2}  M^{\boldsymbol{\theta}}_2 X^{s_1}_2  Z^{s_1}_3  M^{\boldsymbol{\theta}}_1E_{3,4}  E_{2,3} E_{1,2}\\
    & \Rightarrow  X^{s_3}_4Z^{s_2}_4 \tensor*[_{s_1}]{[M^{\boldsymbol{\psi}}_3]}{^{s_2}}   [M^{\boldsymbol{\phi}}_2]^{s_1}    M^{\boldsymbol{\theta}}_1E_{3,4}  E_{2,3} E_{1,2}
  \end{align*}
  This pattern is now standardised. It is clear that it now consists first of entangling commands, then $X$-dependent measurements and finally corrections on the output qudit. In this case as there are no Clifford operators the Pauli simplification stage changes nothing. Signal-shifting is applied which results in the transformation
  \begin{align*}  \mathfrak{p}^{s} & \Rightarrow   X^{s_3-s_1}_4Z^{s_2}_4 [M^{\boldsymbol{\psi}}_3]^{s_2}   [M^{\boldsymbol{\phi}}_2]^{s_1}    M^{\boldsymbol{\theta}}_1E_{3,4}  E_{2,3} E_{1,2}
  \end{align*}
This sequence is then completely standardised. Notice that although this procedure has (slightly) reduced the depth of the pattern, all of the measurements are still dependent and have to be performed in sequence. To demonstrate the procedure when some of the gates are Clifford, return to the standardised pattern $\mathfrak{p}^{s}$ and set $\boldsymbol{\psi}=\boldsymbol{p}$. The Pauli simplification procedure obtains the pattern
 \begin{align*}   \tilde{\mathfrak{p}}^{s}  & =  X^{s_3}_4Z^{s_2}_4 \tensor*[_{s_1}]{[M^{\boldsymbol{p}}_3]}{^{s_2}}   [M^{\boldsymbol{\phi}}_2]^{s_1}    M^{\boldsymbol{\theta}}_1E_{3,4}  E_{2,3} E_{1,2} \\
 & \Rightarrow  X^{s_3}_4Z^{s_2}_4 \tensor*[_{s_1+s_2}]{[M^{\boldsymbol{p}}_3]}{}   [M^{\boldsymbol{\phi}}_2]^{s_1}    M^{\boldsymbol{\theta}}_1E_{3,4}  E_{2,3} E_{1,2}\\
& \equiv \tilde{\mathfrak{p}}^{ps} 
  \end{align*}
 Applying signal shifting to this new command sequence then results in the pattern
  \begin{align*} \tilde{\mathfrak{p}}^{ps}   & \Rightarrow  X^{s_3-s_2-s_1}_4Z^{s_2}_4 M^{\boldsymbol{p}}_3   [M^{\boldsymbol{\phi}}_2]^{s_1}    M^{\boldsymbol{\theta}}_1E_{3,4}  E_{2,3} E_{1,2}
  \end{align*}
This pattern is now completely standard. Notice that the Clifford measurement has no dependencies and hence may be implemented in the first round of measurements.
  
  \section{\label{Opt} }
In this appendix it is shown that there are a large range of quantum computational models which cannot have a lower depth complexity than measurement patterns.
\begin{prop}
Consider a quantum model $ \mathfrak{M}= (\mathcal{A},\mathcal{S} ) $ in which the set of allowed operations $\mathcal{A}$ consist only of
\begin{enumerate}
\item Unitary operators in $\textsc{QMNC}_{d}^0$,
\item Destructive measurements of self-adjoint operators $\hat{O}$ acting on any number of qudits with outcomes in $\mathbb{Z}(d)$ such that $U \hat{O} U^{\dagger}$ is diagonal in the conjugate basis for some $U \in \textsc{QMNC}_{d}^0$,
\item Unitary operators $u  \in \textsc{QMNC}_{d}^0$ that are classically controlled by dits calculated from previous measurement outcomes.
\end{enumerate}
and where the set of preparable states $\mathcal{S}$ for the non-input qudits is such that
\begin{enumerate}
 \setcounter{enumi}{3}
\item For each $\ket{\psi} \in \mathcal{S}$, $\ket{\psi} = U \ket{+_0}$ for some $U \in \textsc{QMNC}_{d}^0$.
\end{enumerate}
For any computation $\mathfrak{Q}$ in such a quantum model there exists a measurement pattern $\mathfrak{P}$ that simulates $\mathfrak{Q}$ in a depth of $O(\mathrm{depth}(\mathfrak{Q}))$. 
\end{prop}
This proposition is similar to one proven for qubits by Browne \emph{et al.} (see Ref. \cite{browne2011computational} Theorem 4). The proof is relatively straightforward. The preparation of all non-input qudits in states from $\mathcal{S}$ can be achieved with initial measurement patterns of constant depth from qudits prepared in $\ket{+_0}$ by condition 4 of the proposition. $\mathfrak{Q}$ may be decomposed into $\text{depth}(\mathfrak{Q})$ sub-computations each of unit depth. In each sub-computation there is at most one operation on each qudit. The unitaries in this layer that are not classically controlled may be implemented with constant depth measurement patterns due to condition 1. Each (in general, many-qudit) measurement in the layer may be simulated by first applying the unitary that diagonalises the measurement in the conjugate basis, which may be done with a constant depth measurement pattern by condition 2, and then implementing $M^{\boldsymbol{0}}_i$ commands (a conjugate basis measurements) on each qudit that the measurement acts on. The appropriate measurement outcome of $\mathcal{O}$ associated with the projection onto the resultant conjugate basis state of the qudit(s) can then be calculated from the individual qudit measurement outcome(s). Note that although this is in general different to a measurement of $\mathcal{O}$ (as $\mathcal{O}$ has outcomes in $\mathbb{Z}(d)$ rather than $\mathbb{Z}(d)^k$ where $k$ is the number of measured qudits), as the measured qudits are discarded (the measurement is destructive) these procedures are identical given the assumption that only the dit calculated from the individual measurement outcomes is retained. As the procedure for each measurement in the layer is of constant (quantum) depth and all the measurements in the layer must act on distinct qudits the measurements may be implemented by a constant depth measurement pattern. The classically controlled unitaries may clearly be implemented with a constant depth measurement pattern as they all act on distinct qudits and are of the form $u^{n}$ for $n \in \mathbb{Z}(d)$ with $u$ implementably with a constant depth measurement pattern by condition 3. Therefore, each component in a layer of $\mathfrak{Q}$ may be implemented with a constant depth measurement pattern and as each operation in the layer acts on distinct qudits (and may only depend on outcomes from previous layers) the composite measurement pattern for the entire layer has constant depth. Hence, the total pattern simulating $\mathfrak{Q}$ has a depth of $O(\mathrm{depth}(\mathfrak{Q}))$.
\newline
\indent
Clifford operators, single-qudit measurements, Pauli corrections and preparation in Pauli eigenstates satisfy the constraints of this proposition. Hence this guarantees the statement in the main text. It is clear however that this proposition is more general (for example it may be applied with multi-qudit measurements).

\section*{References}
\bibliographystyle{apsrev}
\bibliography{Bib_Library}

\begin{thebibliography}{56}
\bibitem[{not({\natexlab{a}})}]{noteMBQC_A}
\bibinfo{note}{Two bases are mutual unbiased when the absolute value of the
  overlap between any basis state of one basis with any basis state of the
  other basis is $1/\sqrt{d}$ where $d$ is the dimension of the Hilbert space.}
\bibitem[{not({\natexlab{b}})}]{noteMBQC_B}
\bibinfo{note}{This is well-known for the qubit subcase, and follows from the
  Euler decomposition for a rotation on the Bloch sphere.}
\bibitem[{not({\natexlab{c}})}]{noteADQC1}
\bibinfo{note}{It may seem more natural to define the single-qudit Pauli group
  as the operators generated by multiplication of $X$ and $Z$. A simple
  justification for the additional phases in the even case is that then the
  standard qubit Pauli group is recovered for $d=2$. The reason for these extra
  phases is related to the fact that $XZ$ has order $d$ for odd dimensions but
  order $2d$ for even dimensions. The interested reader is referred to the
  given references.}
\bibitem[{not({\natexlab{d}})}]{noteMBQC_C}
\bibinfo{note}{This is of course assuming that the model does not use
  unreasonable classical resources. It is also the case that the wait time
  needed to do any classical calculations between quantum steps may be of
  importance.}
\bibitem[{not({\natexlab{e}})}]{noteMBQC_D}
\bibinfo{note}{This is without lose of generality as any pattern may be
  completely standardised with no increase in size or depth}.
\end{thebibliography}

 \end{document}